\documentstyle[aps,preprint]{revtex}

\begin{document}

\tightenlines

\newcommand{\cst}{C$_{70} $}
\newcommand{\cs}{C$_{60} $}
\newcommand{\csm}{C$_{60}^{-} $}
\newcommand{\cv}{C$_{20} $}
\newcommand{\cvq}{C$_{24}$}
\newcommand{\cvs}{C$_{26}$}
\newcommand{\cvo}{C$_{28}$}
\newcommand{\cts}{C$_{36}$}
\newcommand{\lsN}{$\lambda/N(0)$}

\title{Room Temperature Organic Superconductor?}

\author{N. Breda$^{1,2}$, R.A. Broglia$^{1,3,4}$,G. Col\`o$^{1,3}$, 
	G. Onida$^{5,6}$, D. Provasi$^{1}$ and E. Vigezzi$^{3}$}
	
\address{$^1$Dipartimento di Fisica, Universit\`a di Milano,
	 Via Celoria 16, I-20133 Milano, Italy}

\address{$^2$INFM, Unit\`a di Milano, Italy}

\address{$^3$INFN, Sezione di Milano, Italy}

\address{$^4$The Niels Bohr Institute, University of Copenhagen,
	 D-2100 Copenhagen, Denmark}

\address{$^5$Dipartimento di Fisica dell' Universit\`a di Roma Tor Vergata,
	 Via della Ricerca Scientifica, I-00133 Roma, Italy}

\address{$^6$INFM, Unit\`a di Roma Tor Vergata, Italy}

\date{\today}
\maketitle

\begin{abstract}
The electron--phonon coupling in fullerene \cvo \ has been calculated from
first principles. The value of the associated coupling constant
\lsN \ is found to be a factor
three larger than that associated with \cs. Assuming similar values of the 
density of levels at the Fermi surface N(0) and of the Coulomb pseudopotential 
$\mu^*$ for \cvo--based solids as those associated with  
alkali doped fullerides A$_3$\cs, one obtains T$_c$(\cvo)$\approx$8T$_c$(\cs)
\end{abstract}

\bigskip
\leftline{PACS: 74.70.Wz, 63.20.Kr, 61.48.+c}

\newpage
The valence properties of small fullerenes \cite{CMila}, in particular of the 
smallest fullerene yet observed \cvo, is a fascinating question at the 
fundamental level as well as in terms of its potential applications for the 
synthesis of new materials~\cite{Kaxi,Guo,Bylan,Dunlop,Peder,Canni}. In 
supersonic cluster beams obtained from laser vaporization, C$_{28}$ is the 
smallest even-numbered cluster, and thus the fullerene displaying the 
largest curvature, which is 
formed with special abundance. In fact, under suitable
conditions, C$_{28}$ is almost as abundant as C$_{60}$~\cite{Guo}. At 
variance with its most famous family member \cs, \cvo \ is expected to form
a covalent crystal (like \cts \ \cite{Gross,Cote,Piskot}), and not a Van der 
Waals solid \cite{Dress}. However, 
similarly to \cs, fullerene \cvo maintains most of 
its intrinsic characteristics when placed inside an infinite crystalline 
lattice~\cite{Kaxi}. The transport properties of the associated metal doped 
fullerides, in particular superconductivity, can thus be calculated in terms 
of the electron--phonon coupling strength $\lambda$ of the isolated molecule, 
and of the density of states of the solid \cite{Schlu,Gunn}. 
In keeping with the fact that curvature--induced hybridization of the 
graphite sheet $\pi$ orbitals, seems to be the mechanism 
explaining (cf.~\cite{Schlu,Gunn,Devos,Crespi} and refs. therein) the large 
increase in T$_c$ in going from graphite intercalated compounds 
(T$_c$$\approx 5$K)~\cite{Bela} to alkali--doped C$_{60}$ fullerides,  
(T$_c$$\approx 30-40$K)~\cite{Tani,Pals,Pals2}, fullerene \cvo \ is a 
promising candidate with which to form a high--T$_c$ material. These 
observations call for an accurate, first--principle investigation of the 
electronic and vibrational properties, as well as of the electron--phonon 
coupling strength of this system. In the present work we present the results 
of such a study, carried out 
within {\it ab--initio} density functional theory (DFT) in the local 
spin density approximation (LSDA). Our findings are that the associated value 
of \lsN \ is a factor $2.5$ and $1.2$ larger than that associated 
with \cs \ \cite{Gunn} and \cts \ \cite{Cote} respectively. Under similar 
assumptions for the density of levels at the Fermi energy N(0) and for the 
Coulomb 
pseudopotential $\mu^*$ as those associated with alkali-doped fullerides
A$_3$C$_{60}$, one will thus expect 
T$_c$(\cvo)$\approx$8T$_c$(\cs), opening the possibility for \cvo--based 
fullerides which are superconducting at, or close to, room temperature.

The equilibrium geometry of \cvo \ obtained in the present calculation is 
similar to that proposed by Kroto and co--workers \cite{Kroto}, and has the 
full T$_d$ point group symmetry. All atoms are three fold coordinated, 
arranged in 12 pentagons and 4 hexagons. The large ratio of pentagons to 
hexagons makes 
the orbital hybridization in \cvo \ more of sp$^3$ type rather than sp$^2$,
the typical bonding of graphite and \cs. The sp$^3$--like hybridization 
is responsible for a series of remarkable properties displayed by small 
fullerenes in general and by \cvo \ in particular.
Some of these properties are : a) the presence of dangling bonds, which 
renders \cvo \ a strongly reactive molecule, b) the fact that \cvo \ can be 
effectively stabilized (becoming a closed shell system displaying a large 
HOMO--LUMO energy gap) by passivating the four tetrahedral vertices either 
from the outside (\cvo H$_4$) or from the inside (U@\cvo)~\cite{Guo}. It also 
displays a number of hidden valences: in fact, \cvo H$_{10}$, \cvo H$_{16}$,  
\cvo H$_{22}$ and  \cvo H$_{28}$ are essentially as stable as  \cvo H$_{4}$ 
(all displaying HOMO--LUMO energy gap of the order of $1.5$ eV) 
\cite{CMila}, in keeping with the validity of the free--electron picture of 
$\pi$--electrons which includes, as a particular case, the tetravalent chemist 
picture,  
c) while typical values of the matrix elements of the deformation potential 
involving the LUMO state
range between 10--100 meV, the large number of phonons which couple to the 
LUMO state produces
a total electron--phonon matrix element of the order of 1 eV (cf. Table 1), as 
large as the 
Coulomb repulsion between 
two electrons in C$_{28}$. This result (remember that the corresponding 
electron--phonon matrix element is $\sim 0.1$ eV and the typical Coulomb 
repulsion is 
$\sim 0.5-1$ eV for \cs \ \cite{Gunn}) testifies to the fact that one should 
expect unusual properties for both the normal and the superconducting state 
of \cvo--based fullerides, where the criticisms leveled off against 
standard theories 
of high T$_c$ of fullerenes (cf. e.g. 
refs. \cite{Gunn,Ander,Pietro,Pietro2,Grima} and refs. therein) will be much 
in place. 

In Fig 1(a), we report the electronic structure of \cvo \ ccomputed
within the Local Spin Density approximation, as obtained from a  
Car--Parrinello~\cite{Car} molecular dynamics scheme~\cite{Hutter,note_tec}. 
Near the Fermi level we find three electrons in a $t_2$ orbital,
and one in a $a_1$ orbital, all with the same spin, in agreement with the 
results of \cite{Guo}. The situation is not altered, aside from a slight
removal of the degeneracy, when the negative anion, \cvo$^-$, is considered 
(see Fig. 1(b)). In this case, the additional electron goes into the $t_2$ 
state, and has a spin opposite to that of the four valence electrons of 
neutral \cvo.

The wavenumbers, symmetries, and zero-point amplitudes 
of the phonons of \cvo \  
are displayed in 
Table 1, together with the matrix elements of the deformation potential 
defining the electron--phonon coupling with the LUMO state. 
The total matrix element summed over all phonons is equal to $710$ meV. 
The partial electron--phonon coupling constants $\lambda_{\alpha}/N(0)$,
also shown in Table 1, sum up to $214$ meV. 
This value is a factor $2.5$ larger than that observed in \cs \ \cite{Gunn},
and a factor $1.2$ larger than the value recently predicted for 
\cts \ \cite{Cote}. In 
Fig. 2 we display the values of \lsN \ for \cst, \cs, \cts \ 
and \cvo \ \cite{Cote,Provasi,Gunn3,Breda}, which testify to the central 
role the sp$^3$ curvature induced hybridization has in boosting the strength 
with which electrons couple to phonons in 
fullerenes~\cite{Schlu,Gunn,Devos,Crespi}.

In keeping with the simple estimates of T$_c$ carried out in 
refs.~\cite{Gunn,Cote} for C$_{60}$ and C$_{36}$ based solids, we 
transform the value of \lsN \ of Table 1 into a critical temperature 
by making use of 
McMillan's solution of Eliashberg equations \cite{McMil,Elias}
\begin{equation}
T_c = \frac{\omega_{ln}}{1.2}
	\exp{[-\frac{1.04(1+\lambda)}{\lambda-\mu^*(1+0.62\lambda)}]},
\label{mcmillan}
\end{equation}
where $\omega_{ln}$ is a typical phonon frequency (logarithmic average), 
$\lambda$ is the electron--phonon coupling and $\mu^*$ is the Coulomb 
pseudopotential, describing the effects of the repulsive Coulomb interaction.
Typical values of $\omega_{ln}$ for the fullerenes under discussion is 
$\omega_{ln} \approx 10^3$K (cf. e.g. \cite{Bethu,Wang}). Values of N(0) 
obtained from nuclear magnetic resonance lead to values of $7.2$ and $8.1$ 
states/eV--spin for K$_3$\cs \ and Rb$_3$\cs, respectively (cf. 
ref. \cite{Gunn} and refs. therein). Similar values for N(0) are expected 
for \cts \ \cite{Cote}. 
Making use of these values of N(0) for all C$_n$--based solids ($n$=70,60,36 
and 28), one obtains $0.2\leq \lambda \leq 3$ for the range of values of the 
associated parameter $\lambda$. The other parameter entering 
Eq. (\ref{mcmillan}), namely $\mu^*$ and which is as important as $\lambda$ 
in determining T$_c$ is not accurately known. For \cs, $\mu^*$ is estimated 
to be $\approx 0.25$ \cite{Gunn}. Using this value of $\mu^*$, and 
choosing N(0) so that T$_c$$\approx 19.5$K for \cs, as experimentally
observed for K$_3$\cs \ \cite{Gunn}, one obtains 
T$_c$(\cvo)$\approx$8T$_c$(\cs) and 
T$_c$(\cvo)$\approx$1.3T$_c$(\cts) \cite{note34}.

We conclude that \cvo--fullerene displays such large electron--phonon coupling 
matrix elements as compared to the repulsion between two electrons in the same 
molecule, that it qualifies as a particular promising high~T$_c$ 
superconductor. From this vantage point of view one can only speculate 
concerning the transport properties which a conductor constructed making use 
of C$_{20}$ \cite{note37} as a building block can display. In fact, this 
molecule is made entirely out of 12 pentagons with no hexagons, being the 
smallest fullerene which can exist according to Euler theorem for 
polyhedra, and thus displaying the largest curvature a carbon cage can have. 

Calculations have been performed on the T3E Cray computer at CINECA, Bologna.

\begin{table}
{\caption{Phonon wavenumbers, symmetries and  
zero-point amplitudes ($\Gamma_\alpha\equiv
(\hbar/2M\omega_\alpha)^{1/2}$) (columns 1, 2 and 3) of the
phonons of C$_{28}$ which couple to the LUMO state. In columns 4 and 5 the 
corresponding electron--phonon matrix elements $g_\alpha$ and partial
coupling constants $\lambda_\alpha/N(0)$ are displayed. In the last row we
show the corresponding summed values. }}

\begin{tabular}{ccccc}
$1/\lambda$ [cm$^{-1}$]  & symm.& $\Gamma_\alpha$ $(10^{-3} \AA)$ & 
Matrix element $g_\alpha$ [meV] & $\lambda_\alpha/N(0)$ [meV] \\ \hline
351  & $E$   & 63.3 & 7.9  & 1.0  \\
391  & $T_2$ & 59.9 & 10.7 & 2.4  \\
524  & $T_2$ & 51.8 & 49.7 & 38.0 \\
565  & $A_1$ & 49.9 & 12.9 & 0.8  \\
570  & $E$   & 49.6 & 37.0 & 12.9 \\
607  & $E$   & 48.1 & 55.7 & 27.5 \\
707  & $T_2$ & 44.6 & 42.5 & 20.6 \\
724  & $T_2$ & 44.1 & 42.8 & 20.4 \\
763  & $A_1$ & 42.9 & 46.2 & 7.5  \\
771  & $T_2$ & 42.7 & 12.4 & 1.6  \\
791  & $T_2$ & 42.1 & 0.9  & 0.0  \\
976  & $E$   & 37.9 & 43.6 & 10.5 \\
983  & $T_2$ & 37.8 & 15.2 & 1.9  \\
1093 & $T_2$ & 35.9 & 3.4  & 1.0  \\
1101 & $A_1$ & 35.7 & 45.2 & 50.0 \\
1116 & $E$   & 35.5 & 68.9 & 22.8 \\
1171 & $A_1$ & 34.6 & 6.4  & 0.1  \\
1191 & $T_2$ & 34.3 & 43.6 & 12.9 \\
1220 & $A_1$ & 33.9 & 30.5 & 2.0  \\
1260 & $T_2$ & 33.4 & 21.2 & 2.9  \\
1306 & $E$   & 32.8 & 57.5 & 13.6 \\
1381 & $T_2$ & 31.9 & 6.7  & 0.3  \\
1414 & $E$   & 31.5 & 49.2 & 9.2  \\
     & & Total:& 710 & 214 \\     
\end{tabular}

\end{table}

\newpage

\begin{figure}
Fig. 1. Kohn--Sham levels of the neutral (a) and negatively charged (b)
\cvo \ cluster calculated within the LSD approximation. $\alpha$ and 
$\beta$ label the $z$--projection of the electron spin and arrows represent
the valence electrons.
\end{figure}

\begin{figure}
Fig. 2. Calculated electron-phonon coupling constant $\lambda/N(0)$ for 
C$_{70}$~\cite{Provasi}, C$_{60}$~\cite{Gunn}, C$_{36}$~\cite{Cote}, 
C$_{28}$ (cf. Table I). 
\end{figure}


\begin{thebibliography}{9}

\bibitem{CMila} C. Milani, C. Giambelli, H.E. Roman, F. Alasia, G. Benedek,
		R.A. Broglia, S. Sanguinetti and Y. Yabana,
		Chem. Phys. Lett. {\bf 258} (1996) 554.

\bibitem{Kaxi}	E. Kaxiras, L.M. Zeger, A. Antonelli, Yu--min Juan,
		Phys. Rev. {\bf B 49} (1994) 8446.

\bibitem{Guo}	T. Guo, M.D. Diener, Yan Chai, M.J. Alford, R.E. Haufler, 
                S.M. McClure, T. Ohno, J.H. Weaver, G.E Scuseria, R.E. Smalley,
                Science {\bf 257} (1992) 1661.

\bibitem{Bylan}	D. Bylander and L. Kleinman,
		Phys. Rev. {\bf B 47} (1993) 10967.

\bibitem{Dunlop}B.I. Dunlop, O. H\"{a}berben and N. R\"{o}sch,
		J. Phys. Chem. {\bf 96} (1992) 9095.
	
\bibitem{Peder}	M. Pederson and N. Laouini,
		Phys. Rev. {\bf B 48}, (1993) 2733.

\bibitem{Canni}	A. Canning, G. Galli and J. Kim, Phys. Rev. Lett. 
                {\bf 78} (1997) 4442. 

\bibitem{Gross}	J.C. Grossman, M. Cot\'e, S.G. Louie and M.L. Cohen,
		Chem. Phys. Lett. {\bf 284} (1998) 344.

\bibitem{Cote}	M. Cot\'e, J.C. Grossman, M.L. Cohen, S.G. Louie, 
		Phys. Rev. Lett. {\bf 81} (1998) 697.

\bibitem{Piskot}C. Piskoti, J. Yager and A. Zettl,
		Nature {\bf 373} (1998) 771.
	
\bibitem{Dress}	M.S. Dresselhaus, G. Dresselhaus and P.C. Eklund,
		{\it Science of Fullerenes and Carbon Nanotubes}, 
		Accademic Press, New York (1996).

\bibitem{Schlu}	M. Schl\"{u}ter, M. Lanoo, M. Needels, G.A. Baraff 
		and D. Tomanek, Phys Rev. Lett. {\bf 68}, (1992) 526.

\bibitem{Gunn}	O. Gunnarsson, Rev. Mod. Phys {\bf 69} (1997) 575.

\bibitem{Devos}	A. Devos and M. Lannoo, Phys. Rev. {\bf B 58}, (1998) 8236.

\bibitem{Crespi}V.H. Crespi, Phys. Rev. {\bf B 60}, (1999) 100.
	
\bibitem{Bela}	I.T. Belash, A.D. Bronnikov, O.V. Zharikov and
		A.V. Pal'nichenko, Synth. Mat. {\bf 36}, (1990) 283.

\bibitem{Tani}	K. Tanigaki, T.W. Ebbesen, S. Saito, J. Mizuki,
		J.S. Tsai, Y. Kubo and S. Kuroshima,
		Nature {\bf 352} (1991) 222.

\bibitem{Pals} 	T.T.M. Palstra, O. Zhou, Y. Iwasa, P.E. Sulewski, 
		R.M. Fleming and B.R. Zegarski,
		Solid State Commun. {\bf 93} (1995) 327.

\bibitem{Pals2}	T.T.M. Palstra, A.F. Hebard, R.C. Haddon and 
		P.B. Littlewood, Phys. Rev. {\bf B 50} (1994) 3462.

\bibitem{Kroto}	H. Kroto, Nature {\bf 329} (1987) 529.

\bibitem{Ander}	P. Anderson, Theories of fullerens T$_c$'s which 
		will not work, August 20, 1991 (unpublished).

\bibitem{Pietro}L. Pietronero and S. Str\"{a}ssler, 
		Europhys. Lett. {\bf 18} (1992) 627.
	
\bibitem{Pietro2} L. Pietronero, S. Str\"{a}ssler and G. Grimaldi
		  Phys. Rev. {\bf B 52} (1995) 10516.

\bibitem{Grima}	C. Grimaldi, L. Pietronero and S. Str\"{a}ssler, 
		Phys. Rev. {\bf B 52} (1995) 10530.

\bibitem{Car}	R. Car and M. Parrinello, Phys. Rev. Lett. {\bf 55}
		(1985) 2471.

\bibitem{Hutter}J. Hutter et al., MPI F\"{u}r Festk\"{o}rperforschung,
		Stuttgart, and IBM research, 1990--1997. The code has
                been partially modified to calculate the matrix elements
                of the deformation potential.

\bibitem{note_tec} The whole calculation (i.e., geometry optimization of the 
                   cluster, Kohn-Sham levels, phonons and deformation 
                   potential) has been carried out by setting \cvo \ in a fcc 
                   supercell of 26 a.u. A norm-conserving 
                   Trouiller-Martins~\cite{Tro91,Fuchs} pseudopotential has 
                   been employed in the calculation, with a cutoff of 40 Ry. 

\bibitem{Tro91} N. Troullier and J.L. Martins, Phys. Rev. {\bf B 43} (1991)
                1993.

\bibitem{Fuchs} M. Fuchs, M. Scheffler, Comput. Phys. Commun. {\bf 119} (1999)
                67.
		
\bibitem{Provasi} D. Provasi, N. Breda, R.A. Broglia, G Col\`o,
		  H.E. Roman and G. Onida, Phys. Rev. {\bf B} (in press).  

\bibitem{Gunn3}	O. Gunnarsson, Phys Rev. {\bf B 51} (1995) 3493.

\bibitem{Breda}	N. Breda, R.A. Broglia, G. Col\`o, H.E. Roman, 
		F. Alasia, G. Onida, V. Ponomarev and E. Vigezzi,
		Chem. Phys. Lett. {\bf 286} (1998) 350.

\bibitem{McMil}	W.C. McMillan, Phys. Rev. {\bf 167} (1968) 331.

\bibitem{Elias}	G.M. Eliashberg, Soviet Phys.--JETP {\bf 696} (1960) 896.

\bibitem{Bethu}	D.S. Bethune, G. Meijer, W.C. Tang, H.J. Rosen,
		W.G. Golden, H. Seki, C.A. Brown and M.S. de Vries,
		Chem. Phys. Lett. {\bf 179} (1991) 181.

\bibitem{Wang}	Z.H. Wang, M.S. Dresselhaus, G. Dresselhaus 
		and P.C. Eklund, Phys. Rev. {\bf B 48} (1993) 1681.

\bibitem{note34} Similar values of T$_c$ are obtained using Allen's 
                 solution~\cite{Allen,Allen2} of Eliashberg equations. 

\bibitem{Allen} P.B. Allen, R.C. Dynes, Phys. Rev. {\bf B 12} (1975) 905.

\bibitem{Allen2} P.B. Allen and  B. Mitrovi\v{c}, Solid State Physics,
edited by H. Ehrenreich, F. Seitz, and D. Turnbull, vol.  {\bf 37}, 
(Academic press, New York, 1982), p. 1.

\bibitem{note37} While no clear evidences have been found in carbon cluster
                 beams for a particularly abundant bare C$_{20}$ cluster, the
                 fully hydrogenated C$_{20}$H$_{20}$ molecule, dodecahedrane, 
                 turns out to be stable~\cite{Paq}.

\bibitem{Paq} L.A. Paquette, R.J. Ternansky, D.W. Balogh, G.J. Kentgen, 
              J. Am. Chem. Soc. {\bf 105} (1983) 5446.  

\end{thebibliography}
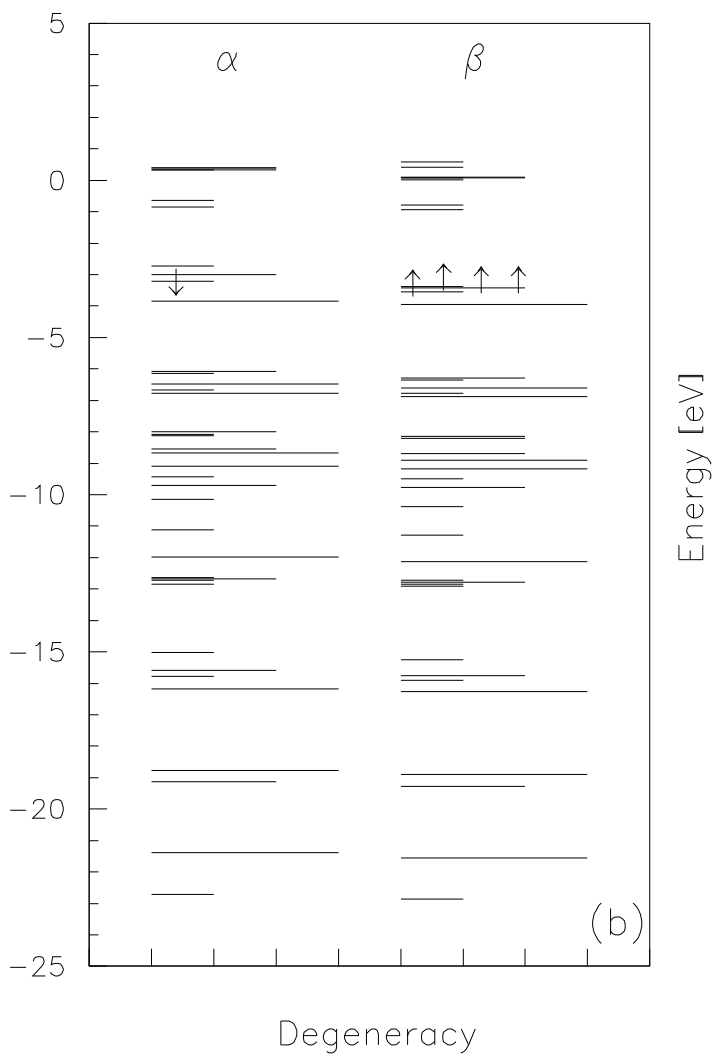
\end{document}